\documentclass[runningheads]{llncs}
\usepackage[T1]{fontenc}
\usepackage{graphicx}
\usepackage{amsmath}
\begin{document}
\title{Synthetic Data for Robust Identification of Typical and Atypical Serotonergic Neurons using Convolutional Neural Networks }
\titlerunning{Synthetic Data for Discriminating Serotonergic Neurons using CNNs}
\author{Daniele Corradetti\inst{1,2} \and
Alessandro Bernardi \and
Renato Corradetti\inst{3} }
\authorrunning{D. Corradetti et al.}

\institute{Grupo de Fisica Matematica (IST), Av. Rovisco Pais, Lisboa, 1049-001, Portugal \and
Departamento de Matematica, Universidade do Algarve, Campus de Gambelas, Faro, 8005-139,Portugal
  \and
Department of Neuroscience, Psychology, Drug Research and Child Health (NEUROFARBA), University of Florence,Viale G. Pieraccini 6,Firenze,50139,Toscana,Italy}

\maketitle              % typeset the header of the contribution
\begin{abstract}
Serotonergic neurons in the raphe nuclei exhibit diverse electrophysiological
properties and functional roles, yet conventional identification methods
rely on restrictive criteria that likely overlook atypical serotonergic
cells. The use of convolutional neural network (CNN) for comprehensive
classification of both typical and atypical serotonergic neurons is
an interesting one, but the key challenge is often given by the limited
experimental data available for training. This study presents a procedure for synthetic data generation that combines smoothed spike waveforms with heterogeneous noise masks from real recordings. This approach expanded the training set while
mitigating overfitting of background noise signatures. CNN models
trained on the augmented dataset achieved high accuracy (96.2\% true
positive rate, 88.8\% true negative rate) on non-homogeneous test
data collected under different experimental conditions than the training,
validation and testing data.

\keywords{Deep Learning Models \and
Serotonergic Neurons \and
Convolutional Neural Networks \and
Synthetic Data \and
Spike Recognition}
\end{abstract}
\section{Introduction}
Serotonergic neurons of the raphe nuclei play an important role in
regulating diverse brain functions and behaviors, including mood,
cognition, sleep, appetite, and pain modulation (Okaty et al., 2019). However,
the serotonergic system is highly heterogeneous, comprised of subpopulations
of neurons with distinct anatomical projections, neurochemistry, physiology,
and functional roles. Elucidating the diversity of serotonergic neurons
is essential to understand how modulatory control of brain states
emerges from serotonergic network dynamics. Traditionally, serotonergic
neurons have been identified in electrophysiology studies based on
\textquotedblleft typical\textquotedblright{} extracellular spiking
characteristics, namely, slow regular firing and long spike duration
(Vandermaelen and Aghajanian, 1983). While useful, these criteria likely
overlook atypical serotonergic neurons, introducing a selection bias
that limits insights into the full diversity of the serotonergic system
(Otaky et al., 2019; Calizo et al., 2011). Deep learning methods like convolutional neural
networks (CNNs) offer a powerful alternative for serotonergic neuron
identification. By learning distinctive spike waveform features, CNNs
can accurately discriminate both typical and atypical serotonergic neurons from non-serotonergic cells
(Corradetti et al., 2024). However, developing robust CNN models for recognizing serotonergic cells from recordings poses practical challenges since large datasets from identified neurons are required for model training, while experimental recordings from serotonergic neurons, identified by methods independent of the signal recordings, are typically limited to hundreds of cells. A first way of overcoming the problem
would be through the use of a naive data augmentation, extracting
short segments as surrogate samples from each recording, thus increasing
by one or two orders of magnitudes the number of labeled samples.
Unfortunately, directly augmenting a limited number of recordings
risks is a huge source of overfitting. A key role in this phenomenon
is represented by the specific noise signature of the experiment which
characterize all the experiments and that is therefore learned by
the model in order rather than meaningful action potential features.
This study addresses the challenges in developing a deep learning model for atypical serotonergic cell recognition using synthetic data generation.
Indeed, this work shows that carefully designed synthetic data augmentation
from limited electrophysiology data can be of use in the development
of deep learning models with high accuracy recognition (Corradetti
et al. 2024). 
The paper is structured as follows. Section 2 discusses the limitations of traditional visual identification methods for serotonergic neurons, while Section 3 examines the problems in applying straightforward deep learning approaches mainly due to the issue of limited experimental data. Section 4 presents a methodology for generating synthetic data to augment the training set. Section 5 describes a practical case study implementing the approach. Finally, Section 6 concludes the paper and discusses future research directions.
\begin{figure}
\begin{centering}
\includegraphics[scale=0.4]{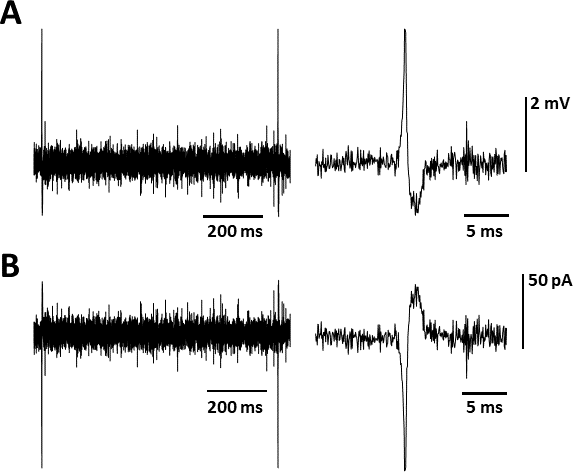}
\par\end{centering}
\caption{\label{fig:idealized}Appearance of extracellularly recorded action potentials according to the recording
arrangement. A. Spikes from a serotonergic neuron recorded in voltage recording configuration,
typically used in vivo or in vitro with sharp microelectrodes. Left trace shows two spikes on a slow
recording timebase. Note the predominance of background noise over the spikes in the recorded
segment. Right trace shows one of the spikes at faster timebase. Note the rapid positive upstroke
followed by a negative phase comprising two subsequent downstrokes. B. Appearance of the
same signals as obtained with the loose-seal patch-clamp current recording configuration used in
the present work.}
\end{figure}
\section{Limitations of Visual Identification Methods }

Serotonergic neurons have historically been identified in electrophysiology
studies using subjective visual criteria centered on \textquotedblleft typical\textquotedblright{}
spiking characteristics. The most commonly applied criteria include
spike shape, spike duration, and regularity of firing (Vandermaelen
and Aghajanian 1983). Idealized  serotonergic neurons exhibit polyphasic spikes with an initial fast deflection followed by one or two slower deflections in the opposite direction (see e.g. Fig. 1). Spike duration, measured from initial rise to second downstroke,
is relatively long, generally >1.2,ms. Firing is regular at slow
rates around 0.5-2.5 Hz. Neurons meeting all these criteria can be classified
as serotonergic with high confidence. However, sole reliance on these
restrictive \textquotedblleft typical characteristics\textquotedblright{}
overlooks the diversity of firing patterns and spike morphologies
across serotonergic subpopulations. Recordings in brain slices from
genetically-identified serotonergic neurons revealed a distribution
of spike durations spanning 0.5 ms to >3 ms, with many neurons exhibiting
spikes narrower than the 1.2 ms criterion (Mlinar et al., 2016; Corradetti
et al., 2024). While the overall distribution skews towards longer durations,
nearly a third of genetically-confirmed serotonergic neurons have
spike widths resembling conventional non-serotonergic neurons. Firing
regularity also varies significantly. Most serotonergic neurons display
regular, slow firing. But burst firing, rhythmic oscillations, and
irregular patterns are observed in certain subpopulations (Hajós et al., 1995; Calizo et al., 2011;  Mlinar et al., 2016). Diversity in firing likely reflects differences in inputs
and intrinsic membrane properties between anatomical groups. Again,
many serotonergic neurons exhibit firing indistinguishable from conventional
non-serotonergic cells. Reliance on narrow spike criteria also overlooks
spike shape variations in serotonergic neurons. While triphasic spikes
are quite typical, spikes range from biphasic to polyphasic (Mlinar
et al., 2016). Non-serotonergic neurons likewise display considerable
variability in spike shape, including broad spikes resembling serotonergic
morphologies.

These data demonstrate the insufficiency of restrictive visual criteria
for comprehensively identifying serotonergic neurons in electrophysiology
recordings. Sole reliance on \textquotedblleft typical characteristics\textquotedblright{}
such as measuring the upstroke/downstroke interval (UDI) may result
not indicative enough for immediate serotonergic neuron identification
and introduces a strong selection bias that likely overlooks much
of the diversity of serotonergic neuron physiology and function. Researchers
adhering strictly to conventional spike criteria may discard neurons
that are genuinely serotonergic but exhibit narrower spike width,
irregular firing, or atypical shape. This precludes studying how functional
differences between serotonergic subpopulations emerge from heterogeneity
in spiking characteristics and intrinsic properties. Conversely, non-serotonergic
neurons with spike width and shape similar to conventional serotonergic
criteria may be erroneously classified without additional verification.
Finally, it would be highly beneficial for research groups to have a model capable of recognising serotonergic cells with high accuracy and in a fraction of a second from a few spike events while the experiment is still ongoing. Such script can be easily
done once a specific deep-learning model is developed from a reasonable
amount of collected data and tailored to the experimental set-up of
the research group. Indeed, the inference time needed for a model
similar to that presented here is of a few millisecond and with an
accuracy definitely higher than that of visual discrimination, thus
suitable for real-life experiments.

\section{Problems with the Deep Learning Methods }

While deep learning methods like CNNs offer immense potential for
serotonergic neuron identification, developing robust models from
electrophysiology data poses a few practical challenges. A fundamental
issue is the limited number of experimental recordings available.
Despite access to an extensive serotonergic cell recording database (Mlinar et al., 2016), the total number of recorded cells was only in the order of a few hundred. The reason for this scarcity
relies on the advanced procedures needed for this type of experiment.
Indeed, since the recognition has to be independent of the recording,
then serotonergic and non-serotonergic neurons must be identified
on the basis of serotonergic system-specific fluorescent protein expression
(serotonergic) or lack of expression (non-serotonergic). The 
procedures needed to obtain the three transgenic mouse lines with
serotonergic system-specific fluorescent protein expression used in
the present work: i) Tph2::SCFP (TSC transgenic mouse line); ii) Pet1-Cre::Rosa26.YFP
(PRY transgenic mouse line); iii) Pet1-Cre::CAG.eGFP (PCG transgenic
mouse line) are explained in detail in (Mlinar et al., 2016, Montalbano
et al., 2015).

Given the scarcity of data, a common tactic would be to expand the
limited data through aggressive augmentation, such as extracting short
segments from longer recordings. A first naive approach in this direction
would be to select segments of a few seconds, in order to obtain multiple
samples of the spike signal along with enough time to assess firing
regularity. However, this naive approach proved to be problematic. Indeed,
the core issue is that each cell recording possesses an intrinsic
background noise signature. This signature becomes a distinct marker
that models learn for discriminating between serotonergic and non-serotonergic
cells. Although segments appear distinct, models can exploit minor
noise correlations rather than encoding robust spike features. One
thus may have excellent accuracy on validation and test data that
are derived from the same experimental recordings as the training
set, but dramatically declined performances on data from different
experiments not used in training. Even more problematic is the fact
that the sources of the noise signature include neighboring cell activity
and probe positioning, both of which change in different cells, but
also environmental factors that remain constant throughout the whole
experimental day.

Our studies (Corradetti et al., 2024) yielded to a few important conclusions
and suggestions: First, each deep learning model must be rigorously
evaluated on \emph{non-homogeneous data} that are not only external
to the training, validation and test dataset, but also collected on
different experimental days than those used in the training and thus
with different noise profiles. Therefore, while one can proceed with
data augmentation extracting small segments from the recording, then
blob all of them before randomizing the splitting in training, validation
and test data; one must also preserve a relevant part of the data,
e.g. >15-20\%, for non-homogenous testing being sure of not having
the data with noise signature similar to that used in the training.
Second, the authors found the model overfitting was highly sensitive to the
length of the extracted segments. Although firing frequency is highly
important for visual discrimination, our deep learning models benefited
immensely from limiting clips to just the central 4 ms region surrounding
spikes (Fig. 2). This showed to improve the focus on the spike alone and reduces
the overfitting given by the noise signature. Finally, a very efficient
solution for expanding the training data is given by the generation
of a synthetic data set for which the authors develop a very specific procedure
(see the next section) that combine smoothed spikes signals along
with real noise masks.

\begin{figure}
\begin{centering}
\includegraphics[scale=0.5]{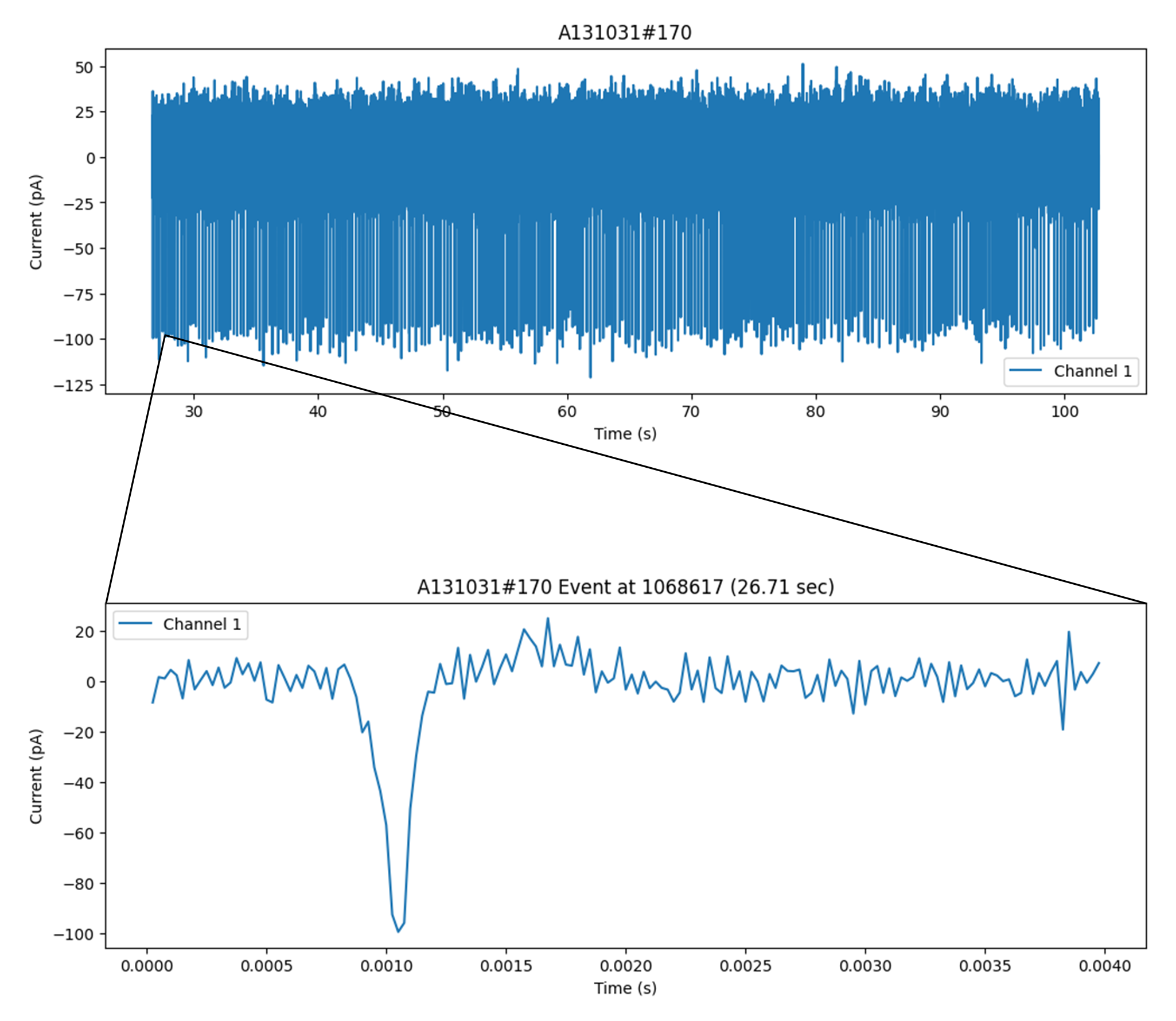}
\par\end{centering}
\centering{}\caption{Example on how single events were isolated and selected. The image
depicts the recording of the serotonergic cell A131031\#170 and the
4 ms event of triggered at point 1068617, i.e. at 26.715 sec.}
\end{figure}

\begin{figure}
\begin{centering}
\includegraphics[scale=1.1]{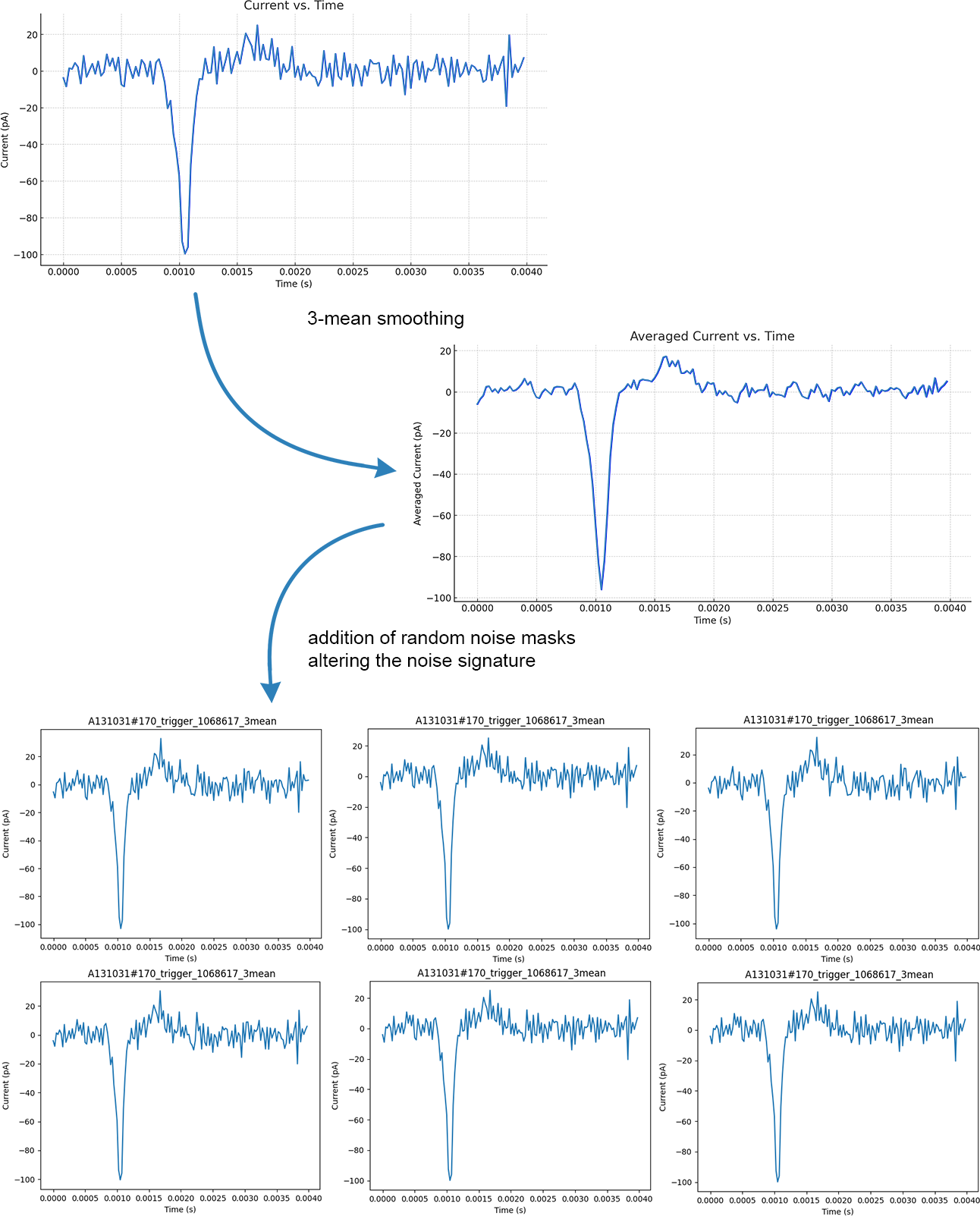}
\par\end{centering}
\caption{\label{fig:SyntheticData}Example of 4 synthetic spikes generated
by the event triggered at 1068617, i.e. 26.715 sec, of the serotonergic
cell A131031\#170. Top trace: the original recording of the event.
The panels report four spike obtained by processing the original trace
with different noise masks (see methods).}
\end{figure}

\section{Generation of Synthetic Data }

The use of synthetic data for deep learning models is increasingly
important nowadays, more so in this case due to the lack of a large
amount of collected data and the difficulty of experiments. In the
generation of synthetic data for emulating spike recordings of serotonergic
cells, some elements are important: firstly, it is paramount to maintain
a spike waveform resulting from a depolarization and repolarization
of the cell consistent with the biological ones; secondly, it is necessary
to avoid imposing a spike morphology that is too uniform and which
does not take into account the biological variability of individual
cells; finally, it is desirable for the signal to contain a plausible
but original and variable noise signature in order to have less sensitive
trained models. The following procedure for generating synthetic data
is developed in order to meet all the previous requirements. Note
that not all requirements point toward the same operational direction.
Indeed, one could think of eliminate from all spikes the background
noise averaging all events (as it is often used for visual discrimination
of serotonergic cells) thus obtaining a theoretical and ideal spike
waveform that one would eventually combine with an \emph{ad hoc} background
noise. That would be an interesting solution that would, nevertheless
compress the biological variability of a cell to a single action potential
waveform. On the other hand, one might want to change arbitrarly the
noise background in order to have very different signatures, but at
the same time the unsupervised application of an arbitrary noise mask
would yield in altering too substantially the spike waveform leading
to unrealistic action potentials.

In our synthetic data generation procedure (Fig. 3), each original training
data sample, i.e., each single event, is smoothed through a simple
moving average (SMA) of range 3. The reason for using a SMA of range
3 is due to the need to combine two requirements: the need to smooth
the original signal from the specific noise of the recording (for
which SMA are a common technique), and the need to maintain the structure
of the signal as mentioned above. Indeed, the rapid depolarization
of the cell is such that the most relevant data of the spike recording
are often condensed in about a dozen of recording points. This means
that considering a SMA with range $n>3$ could undermine the fundamental
information inside the signal, while $n=2$ might not be sufficient
to remove the background noise. A visual inspection of the averaged
signal in \ref{fig:SyntheticData} shows that the bottom of the event
is not altered by applying an SMA of range 3, while a higher ranged
SMA could create a smooth bottom instead of a spike. Concretely, supposing
a $4$ ms sample recorded at 40 kHz, the values of the smoothed sample
$\left\{ y'_{m}\right\} $ with $m\in\left\{ 1,...,160\right\} $
are given as the averages of the values of the original sample $\left\{ y_{m}\right\} $
by
\begin{equation}
\begin{cases}
y'_{1}=\frac{\left(y_{1}+y_{2}\right)}{2},\\
y'_{m}=\frac{\left(y_{m-1}+y{}_{m}+y{}_{m+1}\right)}{3}, & 1<m<160\\
y'_{160}=\frac{\left(y_{159}+y_{160}\right)}{2}.
\end{cases}
\end{equation}
 
\begin{figure}
\includegraphics[scale=0.5]{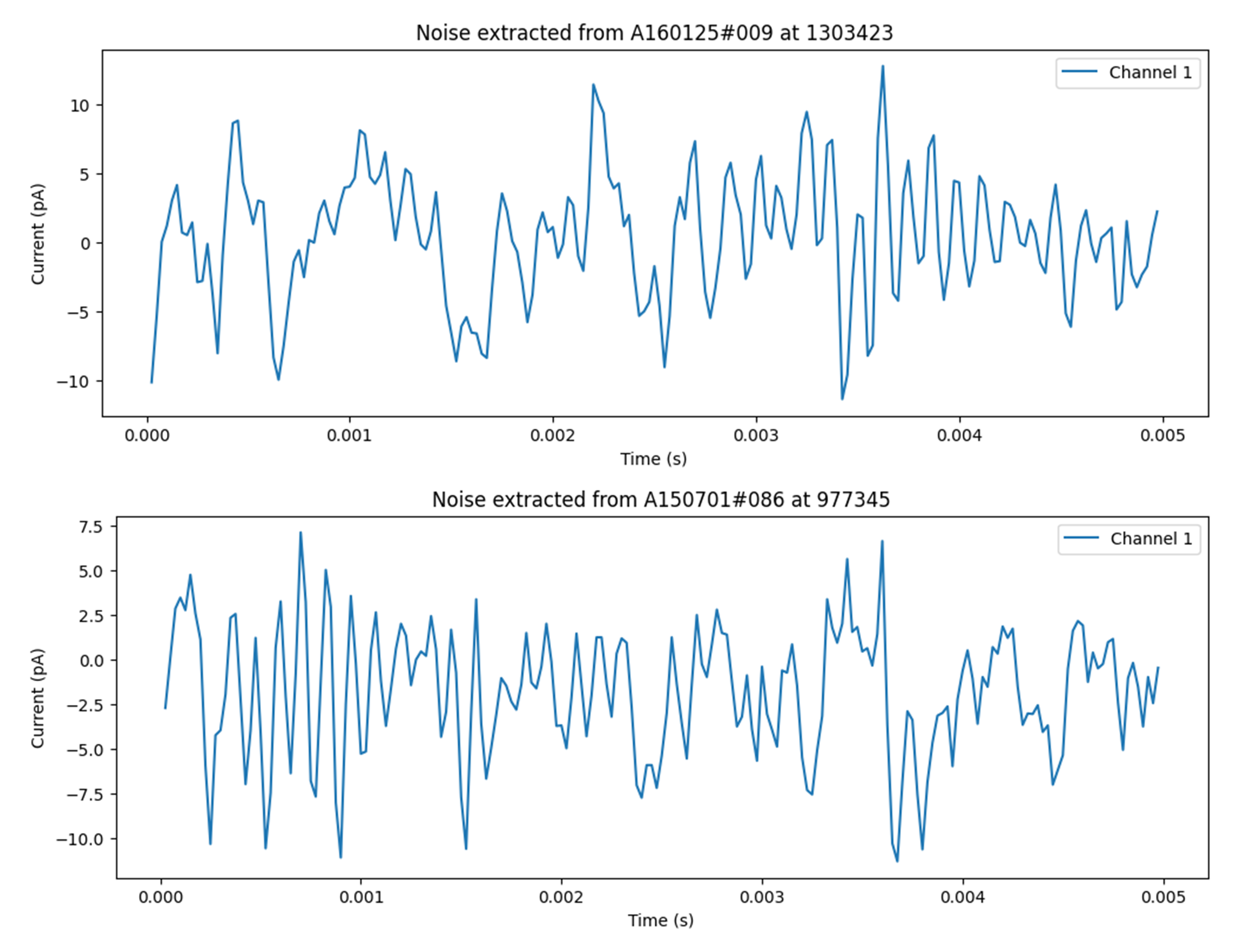}\caption{\label{fig:Noise-masks-collected}Examples of noise masks collected
from the recordings of cell A160125\#009 (\emph{on top}) and
A150701\#086 (\emph{on bottom}).}
\end{figure}

After the previous process has taken place, one has to recombine the
averaged signal $\left\{ y'_{m}\right\} $ with a set of noise masks
$\left\{ n_{m}^{(k)}\right\} $ previously extracted (Fig. 4). From a practical
standpoint, a good way to select such masks is to proceed with an
extraction from real recordings of experiments performed on different
days and under different environmental conditions. A noise mask is
selected extracting a segment immediately before the trigger of an
event, e.g., the noise recording from 5.5 ms to 2.5 ms before the
peak of an event. This ensures the mask contains only background noise
patterns uncorrelated with spike waveform features. Extracting masks
in this manner from multiple heterogeneous recordings provides a diverse
noise set to add variability. Moreover, considering it is quite feasible
to obtain more than $10$ days of recordings, then selecting just
100 noise masks from each day would yield a 1000-fold increase in
data augmentation through synthetic data generation. Indeed, combining
$k=1000$ heterogeneous noise masks enables synthesizing 1000 distinct
versions of each averaged spike. For computational and practical reasons,
the augmentation multiplication factor $R$ is typically constrained
to be considerably smaller than $k=1000$. Nevertheless, the authors suggest
creating a noise mask pool with at least 1000 elements. Then for each
original spike, randomly select $R$ masks from this large pool to
generate the synthetic augmented samples. This approach helps ensure
diversity and minimize correlations in the applied noise patterns. 

The generation of $R$ synthetic data is then obtained from  the values
of smoothed spike $\left\{ y'_{m}\right\} $, that are then added
to the values of $R$ randomly chosen noise mask $\left\{ n_{m}^{(j)}\right\} $
where $j\in\left\{ 1,...,1000\right\} $ is randomly chosen. The final
synthetic sample is thus obtained as the sample $\left\{ y^{(j)}{}_{m}\right\} _{j\in\left\{ 1...R\right\} }$
with
\begin{equation}
y^{(j)}{}_{m}=y'_{m}+\alpha\cdot n_{m}^{(j)},
\end{equation}
where $\alpha\in\left[0.2,0.4\right]$ is a randomly generated ``dumping
coefficient'' experimentally found around $0.3$ to modulate the
noise. The choice of this coefficient requires some clarification.
Indeed, the coefficient dumps the noise intensity to synthesize more
physiologically plausible spike waveforms. First, the background noise
was not completely removed when averaging spikes, just smoothed with
a 3-point average. Thus directly adding the full noise mask would
excessively boost the background noise compared to the original recording.
Moreover, the original noise does not influence all points of the
signal equally, but is more pronounced in slower changing current
regions. Applying the raw noise mask tends to produce unrealistic
spike shapes, e.g. double bottoms. The dumping coefficient $\text{rand}(0.2,0.4)$
was deemed a suitable range by visual inspections by an expert author
with over 30 years of experience on serotonergic spike recordings. 

\section{A Practical Case}

As a practical application of the previous procedure is given by the
model for serotonergic cell recognition now available on GitHub at
\texttt{github.com/ neuraldl/ DLAtypicalSerotoninergicCells.git} .
To implement the model the authors used the following

\paragraph{Original Training Data: }

The original training data for the training, validation and testing
of the models consisted in 43,327 spike samples extracted from 108
serotonergic cells and 45 non-serotonergic cells. More specifically,
the authors extracted 29,773 spikes from serotonergic cells, and 13,554 spikes
from non-serotonergic cells. In all cases, the triggering threshold
of the event was -50 pA and the spike was then sampled 1 ms before
the triggering threshold until 3 ms after (see Fig. 1). Since the
sampling rate of the original recordings was 40 kHz, every spike sample
consists of 160 values. 

\paragraph{Non-homogenous Data:}

The non-homogenous data consisted in 24,616 samples extracted from
55 serotonergic cells and 27 non-serotonergic cells collected in experimental
days not used to obtain the training data, thus with different noise
signature. Again, the triggering threshold of the event was -50 pA
and the spike was then sampled 1 ms before the triggering threshold
until 3 ms after, yielding to 18595 spikes from serotonergic cells
and 6021 from non-serotonergic cells. These data were never part of
the training set, nor validation, nor testing set during the training,
and constituted just an additional independent test for the already
trained model.

\paragraph{Synthetic Data:}

The synthetic data consisted in 12,700,600 spike samples of 160 points
(simulating 4 ms at 40 kHz of sampling) arising from the 43327 original
training data samples. From the original training data recordings
the authors extracted 600 noise masks that constituted the pool for the random
noise selection for obtaining the synthetic data. 

\paragraph{The Architecture of the Neural Network:}

Identifying serotonergic cells is a binary classification task, where cells are categorized as either serotonergic or non-serotonergic. Convolutional neural networks (CNNs) have demonstrated remarkable performance in this domain. Inspired by the structure of the animal visual system, particularly the human brain, CNNs excel at image feature extraction, a crucial aspect of recognition tasks (Liu, 2018). These networks utilize techniques such as feedforward inhibition to mitigate problems like gradient vanishing, thereby enhancing their performance in complex pattern recognition challenges (Liu et al., 2019). 

Considering these advantages, the authors opted for a CNN architecture for the somewhat atypical application of numerical pattern recognition, specifically for analyzing the electrical signals from neuronal cells. This architecture comprises a series of layers typical in image recognition with deep learning using CNNs. The implementation was carried out using the Keras library within TensorFlow 2.15. The network includes a normalization layer to stabilize learning and expedite training, two sets of a 2D convolutional layer with 32 filters each, followed by a max pooling layer with a pool size of (2x1). It also features a flatten layer that connects to a dropout layer and subsequently to dense layers, which have two output units for the binary classification task. The activation function for the convolutional layers is ReLU, while the dense layers utilize the sigmoid function, detailed in Table 1. For training, the authors employed the 'binary crossentropy' loss function, a standard in binary classification tasks, and 'Adam' (Adaptive Moment Estimation) as the optimizer, given its widespread use and effectiveness.
\begin{table}
\centering{}%
\begin{tabular}{ccc}
\emph{Layer (type)} & \emph{Output Shape} & \emph{Param \#}\tabularnewline
\hline 
Layer Normalization & (None, 160, 2, 1) & 320\tabularnewline
Conv2D & (None, 141, 2, 32) & 672\tabularnewline
MaxPooling2D & (None, 70, 2, 32) & 0\tabularnewline
Conv2D & (None, 51, 2, 64) & 41024\tabularnewline
MaxPooling2D & (None, 25, 1, 64) & 0\tabularnewline
Flatten & (None, 1600) & 0\tabularnewline
Dropout & (None, 1600) & 0\tabularnewline
Dense & (None, 2) & 3202\tabularnewline
\hline 
\emph{Total Params} & 45218 & \tabularnewline
\end{tabular}\caption{Schematic of the neural network used with kernel 20. All other models follow the same architectural structure and change only for the dimension of the kernel. The final prediction is given by the consensus of the models with kernel from 20 through 30.}
\end{table}
 A special treatment was devoted to the kernel of the 2D convolutional
layers. Indeed, since the kernel of these layers express the ability
of the convolutional process in enlarging a specific portion of the
pattern, the authors explored a range of possible kernels between 1 to 31.
The training involving > 12M spike samples required a continuous learning
implementation, where the model was trained over 200 training sessions
of 63450 synthetic spike samples. More specifically, for each one
of the 200 training sessions the 63450 synthetic spike samples (33.350
serotonergic and 30.100 non-serotonergic) the authors considered 44.415 spike
for the effective training, 9.517 spike for validation and 9.518 for
test. In each session, all models were trained on 25 epochs with a
batch size of 64. 

To enhance the robustness of the model, instead of selecting a single
kernel and using one model for inference, the authors selected all models with
kernels ranging between 20 and 30 and took the consensus between the
models. This technique ensures more stability in the overall architecture
and is often considered best practice.

\paragraph{Results on Test Data}

Being trained over 12M spikes, i.e. 6,675,300 from serotonergic spikes
and 6,025,300 originated from non-serotonergic spikes, the resulting
model has impressive metrics on the test data. More specifically,
the best training session has a test loss of $1,83\cdot10^{-6}$,
accuracy of $1$, sensitivity at specificity 0.5 of 1, 0 False Positive
and 0 False Negative. However, these results are not deemed significant,
as overfitting not related to recording noise tends to be amplified
in the augmented dataset. 

\paragraph{Results on Non-Homogenous Data }

The most significant outcomes are on non-homogeneous data, i.e., cells
that were not utilized in training and were collected on different
days than the training data. Using this dataset, the synthetic model achieved an accuracy of 0.9375, a sensitivity at specificity 0.5 of 0.8888, an AUC of  0.9255 and an F1-Score of 0.9056.
Even though later sessions have all similar
metrics, the best training session was achieved in session 89 with
the following metrics non-homogenous data (also results on the biological model on the same dataset are reported for comparison).

\medskip{}

\begin{tabular}{c|cccc}
\emph{Model}  & \emph{Accuracy } & \emph{ Sens. at Spec. 0.5}& \emph{AUC }& \emph{F1-Score }\tabularnewline
\hline 
\emph{Biological model} & 0.9125 & 0.8518 & 0.8976 & 0.8679\tabularnewline
\hline 
\emph{Synthetic model}  & 0.9375 & 0.8888 & 0.9255 & 0.9056\tabularnewline
\end{tabular}
 
\medskip{}

A crucial indicator of performance is the model's confusion matrix
specifically a 96.2\% True Positive Rate, 3.7\% False Negative Rate;
88.8\% True Negative Rate, and 11.1\% False Positive Rate. 

\section{Conlusion and Discussion}
This paper presented a methodology for generating synthetic spike data to train deep learning models in identifying typical and atypical serotonergic neurons using smoothed real spike waveforms with diverse noise masks extracted from different real experiments. A practical case study demonstrated the effectiveness of the method, with a CNN model trained on the augmented dataset achieving high accuracy on non-homogeneous test data. While synthetic data have proven effective, the approach may have limitations in fully capturing the diversity of real spiking patterns. Indeed, a special care must be taken during waveform smoothing and noise intensity calibration to preserve key features and avoid creating unrealistic spikes. Moreover it is important to stress out that the method was developed and validated for serotonergic neurons in mice, and its applicability to other species and cell types might require further investigation. 
Despite these limitations, the proposed approach enables the development of robust serotonergic neuron classifiers and opens up to future research (one would like to investigate advanced generative models like GANs and adaptive augmentation strategies). As experimental methods advance and more diverse serotonergic neuron datasets become available, the presented approach can be refined and extended.

\section*{References}
\begin{quote}
Calizo, L. H.; Akanwa, A.; Ma, X.; Pan, Y.; Lemos, J. C.; Craige,
C.; Heemstra, L. A.; Beck, S. G. \emph{Raphe Serotonin Neurons Are
Not Homogenous: Electrophysiological, Morphological and Neurochemical
Evidence}. Neuropharmacology 2011, 61 (3), 524\textminus 543.

Corradetti, D.; Bernardi, A.; Corradetti R.; \emph{Deep Learning Models
for Atypical Serotoninergic Cells Recognition}. bioRxiv 2024.03.03.583157;
doi: https://doi.org/10.1101/2024.03.03.583157

Hajós, M., Gartside, S.E., Villa, A.E., Sharp, T., 1995. \emph{Evidence
for a repetitive (burst) firing pattern in a sub-population of 5-
hydroxytryptamine neurons in the dorsal and median raphe nuclei
of the rat}. Neuroscience 69(1):189-97. doi:10.1016/0306-
4522(95)00227-a.

Mlinar B., Montalbano A., Piszczek L., Gross C., Corradetti R. (2016).
\emph{Firing properties of genetically identified dorsal raphe serotonergic
neurons in brain slices}. Front. Cell Neurosci. 10:195. 10.3389/fncel.2016.00195 

Montalbano, A., Waider,J., Barbieri,M., Baytas,O., Lesch,K.P., Corradetti,R.,
et al (2015). \emph{Cellular resilience: 5-HT neurons in Tph2 (-/-)
mice retain normal firing behaviour despite the lack of brain 5-HT}.
Eur. Neuropsychopharmacol. 25, 2022--2035. doi:10.1016/j.euroneuro.2015.08.021
 
Okaty BW, Commons KG, Dymecki SM. 2019 \emph{Embracing diversity in the 5-HT neuronal system}. Nat Rev Neurosci. 2019 20(7):397-424. doi:
10.1038/s41583-019-0151-3

Liu, Y. (2018). \emph{Feature Extraction and Image Recognition with
Convolutional Neural Networks}. Journal of Physics: Conference Series,
1087. https://doi.org/10.1088/1742-6596/1087/6/062032.

Liu, L., Yang, S., \& Shi, D. (2019). \emph{Advanced Convolutional
Neural Network With Feedforward Inhibition}. 2019 International Conference
on Machine Learning and Cybernetics (ICMLC), 1-5. 

Vandermaelen, CP., \& Aghajanian,G.K. (1983). \emph{Electrophysiological
and pharmacological characterization of serotonergic dorsal raphe
neurons recorded extracellularly andintracellularly in rat brain slices}.
Brain Res. 289, 109--119. doi:10.1016/0006-8993(83)90011-2. 
\end{quote}
\end{document}